**High-precision measurements of krypton and xenon isotopes with a new static-mode Quadrupole Ion Trap Mass Spectrometer**

G. Avice[1], A. Belousov[2], K. A. Farley[1], S. M. Madzunkov[2], J. Simcic[2], D. Nikolić[2], M. R. Darrach[2], C. Sotin[2]

1: California Institute of Technology, GPS division, 1200 E. California Blvd, Pasadena, CA 91125, USA, gavice@caltech.edu, avice.guillaume@gmail.com.

2: Jet Propulsion Laboratory, California Institute of Technology, 4800 Oak Grove Dr., Pasadena, CA 91109, USA.

**9 figures, 4 tables, 9825 words in total**

Significance to JAAS: This study demonstrates the ability of a new Quadrupole Ion Trapp Mass Spectrometer developed at the Jet Propulsion Laboratory (Caltech/NASA) to measure the abundance and isotopic composition of small amounts of noble gases in terrestrial and extraterrestrial samples with high precisions. The QITMS is a compact, versatile and accessible instrument. It is also able to measure gas in static vacuum for extended periods of time (up to 48h). For all these reasons, it could represent a new  promising analytical tool in noble gas geo/cosmo-chemistry and geochemistry in general.

**Abstract**

Measuring the abundance and isotopic composition of noble gases in planetary atmospheres can answer fundamental questions in cosmochemistry and comparative planetology. However, noble gases are rare elements, a feature making their measurement challenging even on Earth. Furthermore, in space applications, power consumption, volume and mass constraints on spacecraft instrument accommodations require the development of compact innovative instruments able to meet the engineering requirements of the mission while still meeting the science requirements. Here we demonstrate the ability of the quadrupole ion trap mass spectrometer (QITMS) developed at the Jet Propulsion Laboratory (Caltech, Pasadena) to measure low quantities of heavy noble gases (Kr, Xe) in static operating mode and in the absence of a buffer gas such as helium. The sensitivity reaches $10^{13}$ cps $Torr^{-1}$ (about $10^{11}$ cps/Pa) of gas (Kr or Xe). The instrument is able to measure gas in static mode for extended periods of time (up to 48 h) enabling the acquisition of thousands of isotope ratios per measurement. Errors on isotope ratios follow predictions of the counting statistics and the instrument provides reproducible results over several days of measurements. For example, $1.7\times10^{-10}$ Torr ($2.3\times10^{-8}$ Pa) of Kr measured continuously for 7 hours yielded a 0.6 ‰ precision on the $^{86}Kr/^{84}Kr$ ratio. Measurements of terrestrial and extraterrestrial samples reproduce values from the literature. A compact instrument based upon the QITMS design would have a sensitivity high enough to reach the precision on isotope ratios (*e.g.* better than 1% for $^{129,131-136}Xe/^{130}Xe$ ratios) necessary for a scientific payload measuring noble gases collected in the Venus atmosphere.

## Introduction

Even if they account for only a minor fraction of the total mass of a planet, atmospheres play a key role in sustaining a planet's habitability[1]. It is thus a major scientific goal to understand their origin and evolution. In this context, measuring abundances and isotope compositions of volatile elements in objects of the solar system permits answers to fundamental questions:(i) the timing and composition of the mixture of volatile elements delivered to terrestrial planets; (ii) the evolution of planetary atmospheres; (iii) the timing and extent of magmatic outgassing from silicate portions of planetary interiors to their atmospheres etc. Studying noble gases is particularly useful here since these elements are inert and thus are only affected by physical processes such as meteoritic/cometary bombardments, thermal and non-thermal atmospheric escape processes, and magmatic outgassing. Furthermore, some of the noble gas isotopes (e.g. $^{40}$Ar, $^{129}$Xe, $^{131-136}$Xe) are produced by radioactive nuclides having half-lives spanning the entire age of the solar system ($t_{1/2}(^{129}I)$=15.7 Ma, $t_{1/2}(^{244}Pu)$=82 Ma , $t_{1/2}(^{40}K)$=1.25 Ga, $t_{1/2}(^{238}U)$=4.47 Ga ). Measuring excesses of these isotopes relative to the base isotope composition permits constraints on the physical processes listed above[2].

Classical Earth-based studies of noble gases in terrestrial or extraterrestrial samples employ magnetic sector mass spectrometers (see the review by Wieler, ref. 3). Some recent developments also involve laser resonance ionization coupled to a time-of-flight technique[4-6]. Sector instruments are often too large and/or consume too much power to be carried by space missions even if some exceptions exist, see for example the mass spectrometer onboard the Pioneer mission to Venus[7]. Laser resonance ionization is also challenging since it involves a complex laser setup and focusing the target gas on a cold spot cooled cryogenically. Recent space missions employ other types of mass spectrometers. Recently, the linear quadrupole mass spectrometer (QMS) of the Sample Analysis at Mars (SAM) instrument suite on-board the Curiosity rover on Mars (MSL mission, NASA) provided results on the isotopic composition of Ar, Kr and Xe in the Mars atmosphere[8,9] and also demonstrated its ability to measure radiogenic ($^{40}Ar$) and cosmogenic ($^{3}He$-$^{21}Ne$-$^{36}Ar$) ages of samples collected from the surface of Mars[10]. Conrad et al. (ref. 9) demonstrated that Mars atmospheric xenon is derived from Solar Xe (SW-Xe) by mass-independent fractionation. The Rosetta ESA mission around the 67P/Churyumov-Gerasimenko comet recently provided results on the elemental and isotope compositions of noble gases in this comet. For example, the double focusing mass spectrometer from the ROSINA instrument suite[11] demonstrated that cometary Xe is depleted in r-process $^{134}Xe$ and $^{136}Xe$ isotopes[12] relative to the Solar composition revealed by the Genesis mission[13]. These examples demonstrate that noble gas geo-/cosmochemistry play an important role in our understanding of the formation and evolution of our solar system. In this context, collecting data on the abundance and isotope composition of noble gases in extraterrestrial atmospheres is a high priority[3]. For example, the composition of the atmosphere of Venus is still a missing piece of the noble gas puzzle[14]. Recent concepts for compact missions sampling the Venus atmosphere below the homopause propose to measure noble gases with an instrument based on the JPL's QITMS (ref. 15).

In this study, we demonstrate that the QITMS developed at JPL is able to detect low abundances of Kr and Xe and to measure the isotope composition of these two noble gases with a precision high enough to answer fundamental questions in cosmochemistry. A previous study demonstrated the ability of the JPL QITMS to measure the isotopic composition of various mixtures of Xe isotopes at a precision better than 0.7‰ (ref. 16). However, these experiments were conducted in dynamic mode with the analytical system connected to a turbo-molecular pump and in presence of a high partial pressure of Xe of about $10^{-6}$ Torr ($10^{-4}$ Pa), two conditions far from natural abundances and classical modes of operation for measuring noble gases. The goal of the present study was to demonstrate the ability of the QITMS to detect, in static mode, much lower abundances of Kr and Xe. Partial pressures were $1.8\text{x}10^{-10}$ Torr ($2.4\times10^{-8}$ Pa) for Kr and $1.3\times10^{-11}$ Torr ($1.7\times10^{-9}$ Pa) for Xe, within about $10^{-6}$ Torr ($10^{-4}$ Pa) of background gases. Furthermore, the QITMS was able to measure, within 1h, the isotopic composition of these two gases in natural samples with a precision good enough to answer scientific questions in noble gas geo-/cosmochemistry and to compete with other instruments in space applications.

## Descriptions of the instrument, calibrating gas and samples

### Instrument

Madzunkov & Nikolić (ref. 16) provided a detailed description of the JPL-QITMS and only the major features and recent modifications are reported in this section. The JPL-QITMS mainly consists of three hyperbolic titanium electrodes (Fig. 1a), corresponding to the classical base design of an ion trap[17]. In the present mode of operation, the top and bottom electrodes (end caps) are grounded and the central ring electrode is powered by a variable high-voltage radiofrequency (*rf*) signal (*rf* frequency =1 MHz, *rf* amplitude ≤ 1.4 kV). A new feature, different from the earlier design, is the use of an electron ionization source mounted on the QITMS ring electrode, providing "side ionization". Neutral gas atoms are ionized within the nappes of a double cone placed apex to apex along the axis of the QITMS. The side-mounted electron gun is composed of a tantalum-disc cathode emitter with an Einzel lens arrangement to focus the electrons through the aperture in the QITMS ring electrode and into the center of the trapping volume. The Einzel lens comprises three electrodes: i) the anode electrode, held at -120 V to prevent electrons from entering the trap during ejection and pulsed to ground potential during ionization, ii) the focus electrode used to tune the transmission of the electron beam; iii) the exit electrode, permanently grounded. Compared to axial ionization scheme[16] for which the electron gun is mounted on one of the endcaps, side-ionization results in a higher number of ions being created near the center of the ion trap, consequently improving the sensitivity of the instrument. Resolution is also higher in the case of side ionization due to the absence of the axially elongated portion of the initial ion cloud. Finally, an electron gun mounted on QITMS ring electrode also offers the possibility to mount two detectors, one on each end cap electrode, with an expectation of roughly doubling the sensitivity (this feature has not yet been implemented). Another modification of the version used in this study is that the ceramic spacers originally in the form of a continuous ring providing electrical insulation between the ring electrode and each end cap electrode are now cut out in 4 distinct pieces. The resulting space (≈ 1 cm wide and several millimeters high (Fig. 1a)) increases the conductance between the volume inside and outside the QITMS, ensuring nearly instantaneous pressure equilibration with the static vacuum chamber. High precision measurements, especially for noble gases in static mode, requires a very high level of cleanliness since contaminants are not pumped away but rather accumulate in the instrument during the measurements. To reduce residual gas levels, all connecting wires previously covered by polyimide material (kapton) have been replaced by bare copper wires electrically insulated by high purity alumina ceramics. Similar to magnetic sector instruments, the new version of the instrument can be baked up to 300 °C, a prerequisite to ensure elimination of residual hydrocarbons often present after venting an instrument[3].

After ionization by the electron beam and trapping by the *rf* electric field, ions are ejected in both directions along the z-axis via a *rf* amplitude ramp applied to the ring electrode (Fig. 1b). Ions are detected with a MAGNUM Channeltron electron multiplier operated in pulse-counting mode. The whole system is mounted on a 6" Conflat flange and placed in a vacuum chamber pumped via a turbomolecular pump (dynamic pressure lower than $10^{-9}$ Torr or $10^{-7}$ Pa) when no gas is analyzed. During a measurement, the chamber is isolated from the turbopump (static mode) and two getters, a SAES CapaciTorr D50 in the purification line and a SAES NP10 in the ion trap vacuum chamber, purify the residual gas and reduce the partial pressure of hydrogen (static pressure in the $10^{-6}$-$10^{-7}$ Pa range). In a second set of experiments, a SAES GP50 getter was placed in the vacuum chamber and the NP10 in the purification line, further improving the purification of residual species (see ref. 3 for a general description of gas purification techniques for noble gas analysis).

**Calibrating gas**

Xenon and krypton abundances in the gas mixture used as a standard in this study have been calibrated in the noble gas laboratory at the California Institute of Technology (Pasadena, USA) using a magnetic sector instrument (Helix SFT mass spectrometer). First, the abundance of helium in a tank of air sampled in Pasadena (California, USA) has been calibrated against the standard used to determine He abundances in geological samples (*e.g.* ref. 18). Abundances of Kr and Xe in the air tank were then deduced from elemental ratios of noble gases in the Earth's atmosphere normalized to He (ref. 2). Kr and Xe fractions from a shot of the same tank were separated from the other gases via condensation on the inner surface of a quartz tube immersed in liquid nitrogen. Kr and Xe were then measured on the magnetic sector instrument in order to determine the sensitivity of the instrument for Xe and Kr. Aliquots from the mixed Xe/Kr cylinder were then measured with the same purification/dilution procedure to determine the abundances of Kr ($2.43 \pm 0.03 \times 10^{-14}$ mol) and Xe ($8.3 \pm 0.5 \times 10^{-16}$ mol) delivered in each aliquot. The isotopic compositions of Kr and Xe in the mixed Xe/Kr cylinder were checked to be atmospheric by measuring geological samples containing modern atmospheric gases.

The QITMS was housed in an ultra-high vacuum chamber with a volume between 4.2 and 4.5 liters. The chamber was constructed from standard vacuum hardware and thus not been optimized for minimum volume (maximum sensitivity). This volume estimate coupled with the gas volume delivered in an aliquot of standard allows us to discuss the sensitivity of the instrument in counts per second per Torr of gas inside the instrument chamber (cps $Torr^{-1}$) where seconds are in wall-clock time.

**Samples and gas extraction techniques**

Gases from chemical residues of a carbonaceous chondrite meteorite and from a zircon ($ZrSiO_4$) megacryst were extracted and purified in the noble gas laboratory of the GPS division at Caltech and measured with the ion trap at the Jet Propulsion Laboratory. The chemical residues are carbon rich materials recovered after acid dissolution (HF/HCl) of a powdered sample of the Allende CV3 carbonaceous chondritic meteorite, and were previously prepared[19]. HF/HCl residues from carbonaceous chondrites typically carry high concentrations ($8 \times 10^{-12}$ mol $g^{-1}$) of Kr and Xe with an isotopic composition distinct from the Earth's atmosphere, the so-called phase Q (refs. 20,21). Gases contained in this residue were extracted by step-heating with a diode laser ($\lambda$=793 nm) at Caltech. In the absence of international standards with a known isotopic composition of Kr and Xe, Allende chemical residues are samples of choice since they carry a well-studied noble gas component with systematic isotope ratios of Kr and Xe (*e.g.* review by ref. 2) that differ from those the Earth's atmosphere which is a classical source of contamination during noble gas measurements[3].

The zircon measured in this study is from high-grade metamorphic terrains in northern Sri Lanka. The age of metamorphic resetting of this terrain is ~600 Ma (ref. 22). Zircons typically contain high concentration of uranium (1000 ppm) and as a result Xe and Kr isotopes produced by spontaneous fission of $^{238}$U (ref. 23) have been accumulating in the zircon for the last ~600

Ma. Gas extracted from these zircon samples is thus a good standard to check the accuracy of the instrument since the fission pattern of $^{238}U$ for Kr and Xe isotopes are well defined (ref. 23). Gas from 900 mg of a single zircon megacryst was extracted at Caltech in a resistance furnace in the presence of 4.5 g of lithium tetraborate ($Li_2B_4O_7$), a chemical reagent used to decrease the melting temperature of silicate minerals. The evolved gas was expanded into a vacuum bottle later isolated from the purification line by two vacuum valves.

Purification of the gas and separation of a pure Kr/Xe fraction were necessary to eliminate hydrocarbons and other impurities released upon heating of the meteorite residue or, in the case of the zircon, to eliminate large quantities of radiogenic He. After laser or furnace extraction, evolved gas was first expanded for 50 min (25 min at 700 °C and 25 min at room temperature) into a stainless steel tube containing Ti-sponge getter material and silver wool. Kr and Xe were then trapped on the walls of a quartz tube immersed in liquid nitrogen (77 K) for 40 min. The non-trapped fraction was pumped out. The quartz tube containing the Kr/Xe fraction was then isolated from the purification line by two vacuum valves and transported to the Jet Propulsion Laboratory. For each extraction, the two valves between the quartz tube and the purification line allowed us to analyze several aliquots. During the experiments, an aliquot of the second purified Kr/Xe fraction of zircon gas (ZIR2) was measured and then mixed with a shot from the standard gas of atmospheric isotopic composition and was called "spiked zircon gas". All sample aliquots were purified using the same procedure as standard gas aliquots prior to inlet into the ion trap volume.

**Measurement cycle**

Although the QITMS *rf* frequency (1 MHz) used in this study allows measurement of light noble gases such as Ne and Ar by changing the trapping voltage, we focused our work on the measurement of Kr and Xe for two main reasons: (i) these two noble gases are in low abundance in the solar system and particularly in planetary atmospheres and are therefore the most challenging species to detect and quantify; (ii) they have numerous stable isotopes some of which are produced by extinct or extant radioactivities. They are thus good tracers of mass-dependent fractionation effects in planetary atmospheres (*e.g.* atmospheric escape) and can also provide time constraints on planetary degassing.

Each aliquot of gas (sample or standard) was first purified by the getter located in the purification line for 10 min before being inlet into the chamber housing the QITMS. During these experiments, the instrument was tuned to maximize the sensitivity for Kr and Xe. The trapping voltage was set to 250 V corresponding to the trapping of ions of *m/z* down to 24 and up to 137. The trapping voltage was high enough to trap $^{136}Xe^{+}$ ions without destabilizing the trajectories of Kr ions. The filament current was set at 1.65 A, corresponding to an emission current of 550 μA. Tuning of the potential applied to the cathode corresponding to the Energy (E) applied to electrons and of the Focus (F) parameter of the Einzel lens was done in two steps: i) choosing an electron energy close to the maximum cross-section for ionization of Kr and Xe by electron impact; ii) tuning the Focus parameter in order to maximize the electron transmission in the trapping region inside the ion trap. An optimal F/E ratio of 0.89 was determined empirically by maximizing the Kr and Xe signals. Finally, the count rate and the voltage applied to the electron multiplier detector were manually scanned in order to find a plateau region where the count rate is high and not susceptible to voltage variations.

One characteristic measurement cycle (50 ms duration) was divided into two distinct intervals (Fig. 1b). The first interval (0 to 22 ms) was dedicated to ionization of gas atoms and trapping of the resultant ions in the center region of the QITMS. During this interval the anode was set to ground potential, allowing electrons to enter the instrument and the trapping voltage amplitude of 250 V was applied to the ring electrode. Two metallic meshes (grids) in front of the detector were powered by +140V and -140V DC voltages to prevent electrons and stray ions from reaching the detector entrance during this interval. The second interval (22 to 50 ms) corresponded to the controlled destabilization of the trapped ions and their ejection from the center of the trap into the detector. During this interval the anode was held at -120V DC potential, preventing electrons from entering the QITMS which would result in further ionization inside the trap volume. The two detector grids were set to ground potential allowing ions to enter the detector region. The ring electrode *rf* amplitude was ramped, typically from 250 V to 1.4 kV, and ions were ejected according to the conventional "mass destabilization mode" [24,25]. All timing and voltage parameters were programmable for the ionization/trapping and ejection phases via a controlling software developed at JPL[16].

## Results

### Mathematical treatment applied to raw data and data reduction

One full mass spectrum ranging from 24 to 137 atomic mass unit (a.m.u) is acquired every 50 ms. It contains $1.6 \times 10^4$ data points (number of channels recorded by the detector) and is hereafter called a "frame". Data counting was performed using commercial National Instruments counting hardware. Unfortunately, the counting board used in this study had a limited bandwidth and recorded only 40-50% of the frames due this limitation. This does not impact our estimates for the sensitivity of the instrument since measurements of signal intensity were done by compiling data by blocks of 20 frames, virtually corresponding to one second of data acquisition. However, isotope data is considered as having been recorded for only 40-50% of the duration of the measurement. Updating the counting hardware would solve the counting issue and is planned in the future. Madzunkov & Nikolić (ref. 16) described displacement of the peaks due to spurious instabilities in the *rf* amplitude, induced for example by temperature variations, and not compensated for by the *rf* generator currently in use. Data presented here were thus re-aligned to account for this effect using the same method as described in detail in ref. 16. This step requires co-adding raw frames into blocks corresponding typically to 2-4 seconds of measurement in order to accumulate a sufficient signal for peak recognition algorithm used during the alignment procedure.

Isotope ratios of standard aliquots are given as raw values *i.e.* without correction for instrumental mass bias (no re-normalization to the isotope composition of atmospheric Xe and Kr). Isotope ratios of Xe and Kr measured in zircon and meteorite samples are corrected for instrumental bias by using the values measured on standard aliquots normalized to the isotopic composition of atmospheric Kr and Xe (ref. 26).

**Sensitivity and signal evolution with time**

Sensitivity defines the quantity of signal (in counts per second, cps) generated by an analytical instrument for a given quantity of gas, here a partial pressure of Kr or Xe in Torr, introduced in the analytical system. In this study, sensitivity was 1.1-1.5x$10^{13}$ cps $Torr^{-1}$ (about $10^{11}$ cps/Pa) for Kr or Xe for the experimental conditions listed in Table 1. The sensitivity is lower than for a terrestrial laboratory magnetic sector instrument by a factor of $\approx 10^3$. For example for an aliquot of calibrating gas, about 1300 cps of $^{84}$Kr are detected with the QITMS whereas a similar gas amount introduced into the Helix SFT would produce a signal of $^{84}$Kr of 2.3x$10^6$ cps. Potential opportunities for increasing the sensitivity of the QITMS are given in the Discussion section.

An example of the evolution of the Kr signal in the QITMS with time is displayed in Fig. 2a and is compared, after normalization, to a Kr signal measured with the Helix SFT magnetic sector instrument. The decay rate (λ) of the signal in the sector instrument is high ($4.4 \times 10^{-4}$ $s^{-1}$) due to the 5 kV differential potential used to accelerate ions and causing their implantation into the metallic surfaces of the instrument[3]. The decay rate of the signal for the QITMS is 100 times lower ($4.5 \times 10^{-6}$ $s^{-1}$) allowing for measurements to proceed for extended periods of time. Examination of the data shows that the decrease in signal in the QITMS, while slow, cannot be solely explained by ion burial on the surfaces of the detector. In the case of Kr, a decrease of 10% of the signal after 6h would correspond to the disappearance of about $10^9$ Kr ions while only $10^7$ ions were counted. Either ions are created and lost for example by implantation on metallic surfaces of the QITMS, or the sensitivity of the instrument declines monotonically during the static-mode analysis. One possible mechanism for such sensitivity loss might be an increase in the partial pressure of a residual gas during the long measurement, for example methane. This residual gas would decrease the temperature of the filament over time and thus its emission. The corresponding decrease in ionization leads to a decrease in sensitivity with time. The possible phenomenon was studied by setting the electron energy at 0 eV ("no ionization" phase) over a 345 min period, followed by setting the electron energy to its nominal 72 eV (thus creating ions). We observed that the resulting Xe and Kr signals at t = 346 min are identical to those measured in a normal experiment where ionization starts immediately. This suggests that sensitivity does not vary with time. Furthermore, measurements of the electron emission coming from the filament remained a constant 550 (±5) μA throughout the experiment. The decrease in signal is therefore almost certainly due to the implantation of ions on metallic surfaces, a phenomenon known to occur for Kr and Xe ionized by electrons with energies as low as 40 eV and able to trap significant amounts of the gases originally present in the gas phase[27]. The signals of Kr and Xe progressively increase when the QITMS is operated in static mode without sample or standard aliquots being introduced in the system. Such an increase corresponds to a memory effect where Kr and Xe are progressively released from metallic surfaces and is in agreement with the hypothesis of ion implantation being responsible for the decay in signal. Trapping of noble gas ions in metals at low energies and mass-dependent fractionation of the trapped fraction (>1%.amu$^{-1}$ for Kr and >0.5%.amu$^{-1}$ for Xe) has previously been reported[27]. In this study however, we did not detect any isotopic fractionation in the free gas phase even after 48h of continuous measurement, when more than 50% of the gas is lost due to trapping in the metal surfaces. However, Ponganis et al. (ref. 27) noted that the trapping phenomenon is a combination of processes involving ion loss in metal by implantation and release of previously trapped ions by sputtering or thermal diffusion[27]. The lack of knowledge on the extent of each of these processes for the QITMS and of their associated isotopic fractionations prevents us to evaluate if, overall, ion implantation could be happening without measurable isotopic fractionation.

**Resolution and peak shape**

The resolution of the QITMS is a complex function of the quality of the generated *rf* signal (amplitude and frequency), the machining tolerance of the hyperbolic electrode surfaces, and the size and position of the ion cloud generated inside the trap[28]. The resolution (m/Δm FWHM) in this study is 246 for Kr and 256 for Xe. However, the trap was tuned for maximum sensitivity, not for maximum resolution. These values are much lower than 600, the typical resolution required to fully resolve Kr and Xe peaks from potential hydrocarbon contributions[29]. This highlights the need for high levels of cleanliness. Kr and Xe peak shapes in this study are distinct from those obtained when operating sector instruments (Fig. 3), in particular for the QITMS we observe: i) a tailing of the peak on the low mass side, a feature typical for ion traps and already recognized in previous studies[16]; ii) weak second and third order peaks on the high mass side of the primary peaks. The latter could be due to an inhomogeneous ion cloud inside the trap or geometric misalignment of the endcap electrodes giving rise to octupole term in the multipole expansion of the trapping potential. However, we observed that peak shapes are reproducible from one analysis to another. Thus, isotope ratios reported in this study were computed by choosing an interval for each peak and summing up the recorded counts within this interval. This leads to slightly inaccurate, but precise, isotope ratios since the influence of the tailing and/or second and third order peaks is not properly taken into account.

**Krypton and xenon isotope ratios of the standard gas**

Isotope ratios of Kr and Xe are reported in Table 2 for 1h experiments and in Table 3 for 7 h experiments.

**Internal reproducibility**

Experiments conducted in this study show that gas aliquots can be measured for extended periods of times (several hours and up to 48h). There is little evolution of the isotope ratios over the course of the measurement (Fig. 2b). There was thus no extrapolation to t=0 s (time of inlet) to compute isotope ratios. Since 20 mass spectra spanning a mass range from 24 to 137 amu are acquired every second, statistical tests can be conducted on the numerous isotopic ratios (N>>100) collected for each measurement (Fig. 4). Isotope ratios computed from data collected during one measurement follow a normal distribution of parameters $\mu$ and $\sigma$, where $\mu$ is the mean isotope ratio and $\sigma$ the standard deviation of the measurement (see Fig. 4a and 4b for the $^{86}Kr/^{84}Kr$ ratio). Since N measurements of the same sample gas are considered, the error discussed here is the standard error of the mean i.e. $\sigma/N^{1/2}$ where N is the number of isotope ratios. The measured error follows closely the error predicted by counting statistics using a Poisson law (Fig. 4c). This demonstrates that the different sources of instabilities for external and internal parameters (temperature, pressure build-up, emission current variations, stability of the *rf* amplitude) do not play a major role in the error budget. The dominant contribution to the error is from counting statistics. Isotope ratios measured for aliquots of the standard significantly differ from those of atmospheric Kr and Xe (Table 2, Figure 5). This is probably due to mass discrimination effects[3] and to the peak integration method used in this study (see above). Thus, similarly to the case of magnetic sector instruments, data must be corrected by normalization to known isotopic ratios of a standard in order to obtain accurate isotope ratios on unknowns.

**External reproducibility: stability of the instrument with time**

Two sets of experiments were conducted in order to characterize the external reproducibility of the instrument. Aliquots of the gas standard were measured for periods of 40min to 1h25min (called 1h experiments hereafter) and 6 to 8h (7h experiments hereafter) over the course of several days (about 10 days for each experiment). Only $^{128-136}Xe/^{132}Xe$ ratios are reported since $^{124}Xe$ and $^{126}Xe$ signals were low and of the same order of magnitude as the hydrocarbon peaks. Examples of the reproducibility of the $^{86}Kr/^{84}Kr$ and $^{129}Xe/^{132}Xe$ ratios from these runs are shown in Fig. 6. For 1h experiments (Table 2), the external reproducibility is 2.2 ‰ and 5.4 ‰ for the $^{86}Kr/^{84}Kr$ and $^{129}Xe/^{132}Xe$ ratios, respectively. Values of the reduced $\chi^2$ are close to 1 for those measurements meaning that the external error is well explained by the internal error of each measurement. However, for 7h measurements (Table 3), reduced $\chi^2$ values are well above 1 (7.5 and 2.6 for $^{86}Kr/^{84}Kr$ and $^{129}Xe/^{132}Xe$ ratios, respectively). Reduced $\chi^2$ remain high even after removing the potential outliers (measurement #3 for the $^{86}$Kr/$^{84}$Kr ratio and measurement #4 for the $^{129}$Xe/$^{132}$Xe ratio). An external source of variability is thus systematically influencing the dataset. This variability might originate from the arbitrary and manual determination of peak limits which is not reproducible between measurements at this level of precision, or from variations arising from the manual gas handling during preparation of the different aliquots. The long term stability (over weeks or months) should be assessed although it requires to leave one version of the instrument untouched for long periods of time. Note that values of isotope ratios for individual measurements and for the datasets are different for 1h and 7h experiments. This is likely due to different values used for the Energy and Focus of the electron gun (Table 1) and highlights the need for standardization by a calibrating gas to get accurate isotope ratios. Preliminary tests confirm that Kr and Xe isotope ratios changes when the Energy and Focus parameters of the electron vary. A future study should focus on exploring the parameter space for Focus and Energy and determining the effects on the instrument sensitivity and mass discrimination.

### Results obtained on natural samples

Results obtained on natural samples are listed in Table 4 and have been corrected for instrument effects using the 1h measurements of the standard gas measured during the same analytical session.

#### Allende

Isotope ratios of Xe and Kr measured in Allende residue samples are listed in Table 4 and the isotopic composition of Xe is displayed in Fig. 7. The average isotopic composition of Kr is close to atmospheric values. This is probably due to the low $^{84}$Kr/$^{132}$Xe ratio in meteoritic residues (0.81, ref. 20,21) compared to the atmospheric ratio (34, ref. 2). The Kr measurements are thus polluted by atmospheric contamination. The average isotopic composition of Xe measured in Allende matches previous values reported in the literature[20,21] for the Q component of carbonaceous chondrites and is clearly distinct from the isotopic composition of atmospheric Xe. The good match of the dataset reported in this study with previous values reported in the literature for Allende confirms that the QITMS provides accurate data after normalization of values obtained on standard aliquots to the isotopic composition of the Earth's atmosphere. The variability of the measurement series is higher than the errors for individual measurements. For example, the $^{130}Xe/^{132}Xe$ varies from 0.1524 ± 0.0051 to 0.170 ± 0.003 resulting in a high MSWD value of 5 and a large error bar for the average value (Fig. 7). This could be due to varying contributions from a hydrocarbon contaminant released by the sample upon heating and not fully removed by the purification procedure used either at Caltech or at JPL before introduction in the QITMS chamber.

**Zircon**

Gas extracted from the zircon yields an isotopic composition of Xe distinct from the composition of the atmosphere (Table 4, Fig. 8). Xe is strongly enriched in $^{136}Xe$, $^{134}Xe$, $^{132}Xe$, $^{131}Xe$ relative to other Xe isotopes confirming the contribution of a fission component. Data fall on a mixing line between air and the pure $^{238}U$ fission end-member (Fig. 8a) with a contribution from $^{238}U$ fission to $^{132}Xe$ of 84.5 % and 91.5 % for ZIR1 and ZIR2 measurements, respectively. This has been computed by using the $^{129}Xe/^{132}Xe$ ratio ($^{129}Xe/^{132}Xe = 0$ for the spontaneous fission of $^{238}U$). In figure 8, the composition of the pure $^{244}Pu$ fission end-member is given only for comparison since the short half-life of $^{244}Pu$ ($T_{1/2}$ = 82 Ma) prevents any incorporation of the parent nuclide at the time of formation (600 Ma ago) of the zircon. The atmospheric contribution could come from several sources: i) air trapped within the zircon megacryst; ii) blank of the resistance furnace or of the lithium tetraborate reagent used to melt the crystal and extract its gas content; iii) air contamination during gas transport from Caltech to the JPL and in the JPL gas preparation line. The latter is probably a major source since the two measurements (ZIR1 and ZIR2) of the same gas extracted at Caltech show different atmospheric contributions. The highest contribution from $^{238}U$ fission to $^{132}Xe$ for ZIR2 is 91.5 % and thus gives a maximum analytical blank contribution of 8.5% of $^{132}Xe$ to the total measured signal for our measurements. When the fission spectrum is corrected for this atmospheric contamination, data plot on the pure $^{238}U$ fission end-member for $^{131}Xe/^{136}Xe$, $^{132}Xe/^{136}Xe$ and $^{134}Xe/^{136}Xe$ ratios (see for example Fig. 8b where the correction is applied to ZIR1). The spiked zircon gas also plots along the mixing line between air and $^{238}U$ fission with a contribution of fission of 27.3 %. For Kr, the $^{86}Kr/^{84}Kr$ ratio is as high as 0.599±0.015 (ZIR2) demonstrating the contribution from a fission contribution (Fig. 8) from $^{238}U_{sf}$. The fission component for Kr is more contaminated by air than for Xe because of a lower abundances of Kr compared to Xe produced by the fission of uranium $((^{136}Xe/^{86}Kr)_{fission}$ = 5-6, ref. 30).

## Discussion

### Potential applications

The QITMS is as a new versatile mass spectrometer able to operate either in dynamic mode[16] or in static mode (this study). It could thus be used as a scientific payload for various space missions such as a probe sampling material in plumes emanating from moons of gas giants[31], mass spectrometry experiment measuring the exosphere of the Moon[32] etc. As part of a large instrument suite (*e.g.* SAM on the Curiosity rover), the QITMS could also measure noble gases released from rocks for geochronology and/or planetary evolution studies. Depending on the application, the QITMS could be accompanied by a gas purification system such as the one present on the Curiosity rover (ref. 33) in order to ensure that only clean gas depleted in major reactive molecules is introduced into the QITMS vacuum chamber. While the QITMS is a promising instrument, one of its current main limitation is a low sensitivity compared to common techniques employing magnetic sector instrument. Future studies could focus on extending the comparison between the two instruments for other types of samples such as volcanic gases. Indeed, despite the low sensitivity, if the quantity of gas is not limited, the low consumption of the QITMS could allow measuring isotope ratios at very high precision. The need for a purification system attached to the instrument is also problematic. Tests are currently carried out to check the ability of the QITMS to remove interfering species (*e.g.* $CO_2$) during ionization via resonant ejection methods[25]. This would relax some of the constraints on the design of a compact purification system since the instrument itself would be part of the cleaning procedure.

The sensitivity and isotope ratio precision determined in this study allow us to evaluate how well the QITMS would perform as an instrument measuring a sample of the Venus atmosphere acquired below the homopause. In a candidate mission scenario[15], an aliquot of the Venus atmosphere would be collected at 110 km at a speed of about 10 km/s leading to a dynamic pressure of 1.58 Torr (210.65 Pa) for the sampled gas. Assuming a QITMS sensitivity of $10^{13}$ cps/Torr (about $10^{11}$ cps/Pa) and abundances of Kr and Xe of 25 and 1 ppb (ref. 34), respectively, a precision better than 1% for all Kr and Xe isotope ratios would be achieved in a single measurement of 3 min with the QITMS. Such values are well within the range of scientific requirements recommended in order to understand significant aspects of the origin and evolution of the Venus atmosphere[14]. Future studies could investigate how the instrument behaves when measuring lighter noble gases.

### Potential future improvements

During this study, the different voltage and timing parameters (Table 1) were optimized for maximum sensitivity leaving little room for further improvements of this aspect of the QITMS. Two more-involved approaches might be pursued in order to improve the sensitivity of the instrument. Firstly, using a custom purpose-built chamber optimized to house the trap with minimum volume would increase the partial pressure of a fixed amount of analyte gas by a factor of at least 5. Secondly, the design of the instrument could be modified by: i) placing a second detector at the exit of the second end-cap electrode, doubling the sensitivity (and also providing redundancy); ii) replacing the Ta-cathode filament by a single $LaB_6$ crystal emitter which provides a tighter electron beam, more electrons would go inside the central trapping region. Finally, an update of the counting hardware would allow to collect 100% of the data generated by the electron multiplier detector.

## Conclusions

Results presented in this study demonstrate that the JPL QITMS is a useful analytical tool to measure noble gases in static mode for extended periods of time (up to 48h per single measurement). The sensitivity of the instrument reaches $10^{13}$ cps/Torr (about $10^{11}$ cps/Pa) of Kr and Xe. Relatively straightforward design modifications could increase the sensitivity to $10^{14}$ cps/Torr ($10^{12}$ cps/Pa). Counting statistics are the dominant contributor to the internal error of single measurements. The external sources of human error introduced during peak integration can be eliminated by developing an automated peak fitting method suitable for low counting statistics. Measurements of terrestrial and extraterrestrial samples demonstrate the ability of the QITMS to produce accurate and precise data. Overall, the instrument represents a new generation of versatile and flexible noble gas mass spectrometer for terrestrial or space applications.

## Acknowledgments

Laurent Rémusat is acknowledged for his advices for acquiring Allende samples originally prepared by S. Epstein and Mark B. Garcia for his handling of the GPS collection at Caltech. Jonathan Treffkorn is thanked for help during the preparation of the experiments conducted at Caltech. This work has been performed at the California Institute of Technology and at the Jet Propulsion Laboratory (JPL), California Institute of Technology (Caltech), under contract to NASA.

# Figures

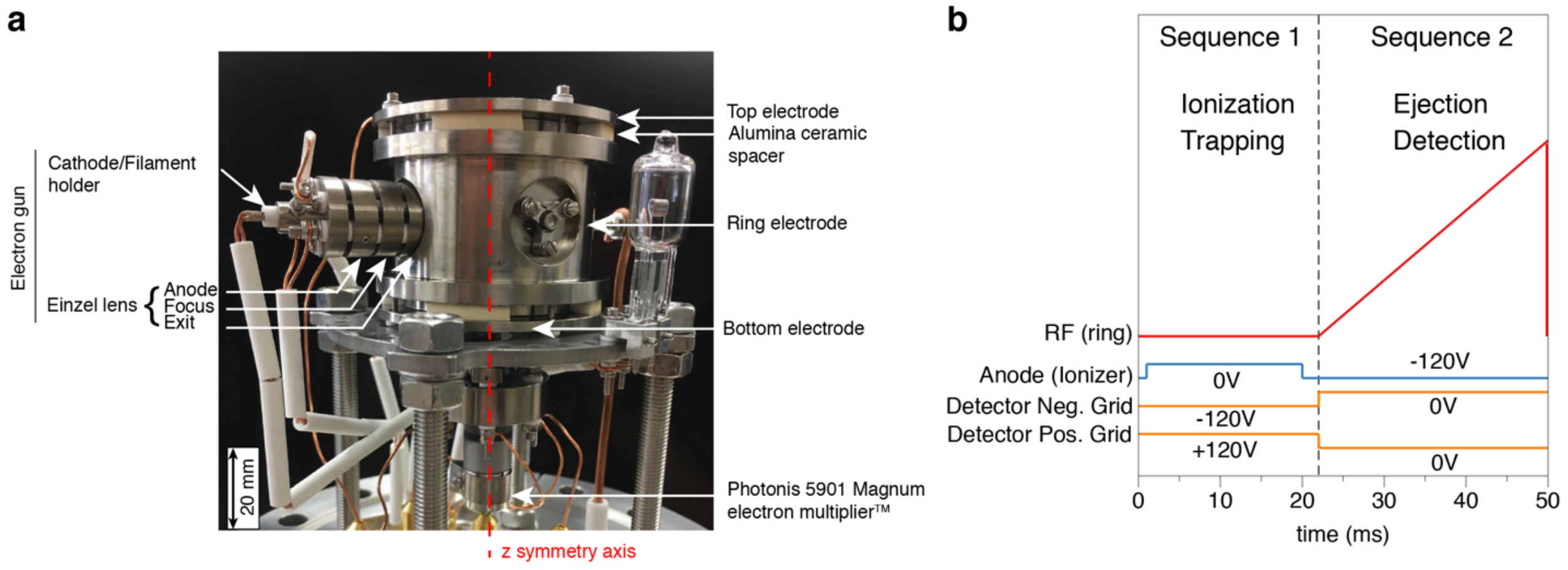

**Figure 1: JPL Quadrupole Ion Trap Mass Spectrometer and its measurement routine.** (a) Picture of the instrument outside the vacuum chamber. (b) Example of a 50 ms measurement routine. Interval 1 (0 to 22 ms) is dedicated to ionization and trapping of ions. Interval 2 (22 to 50 ms) corresponds to the ejection of trapped ion by ramping the voltage applied to the ring electrode and to their detection by the electron multiplier operated in pulse counting mode. All the parameters are adjustable in amplitude and time.

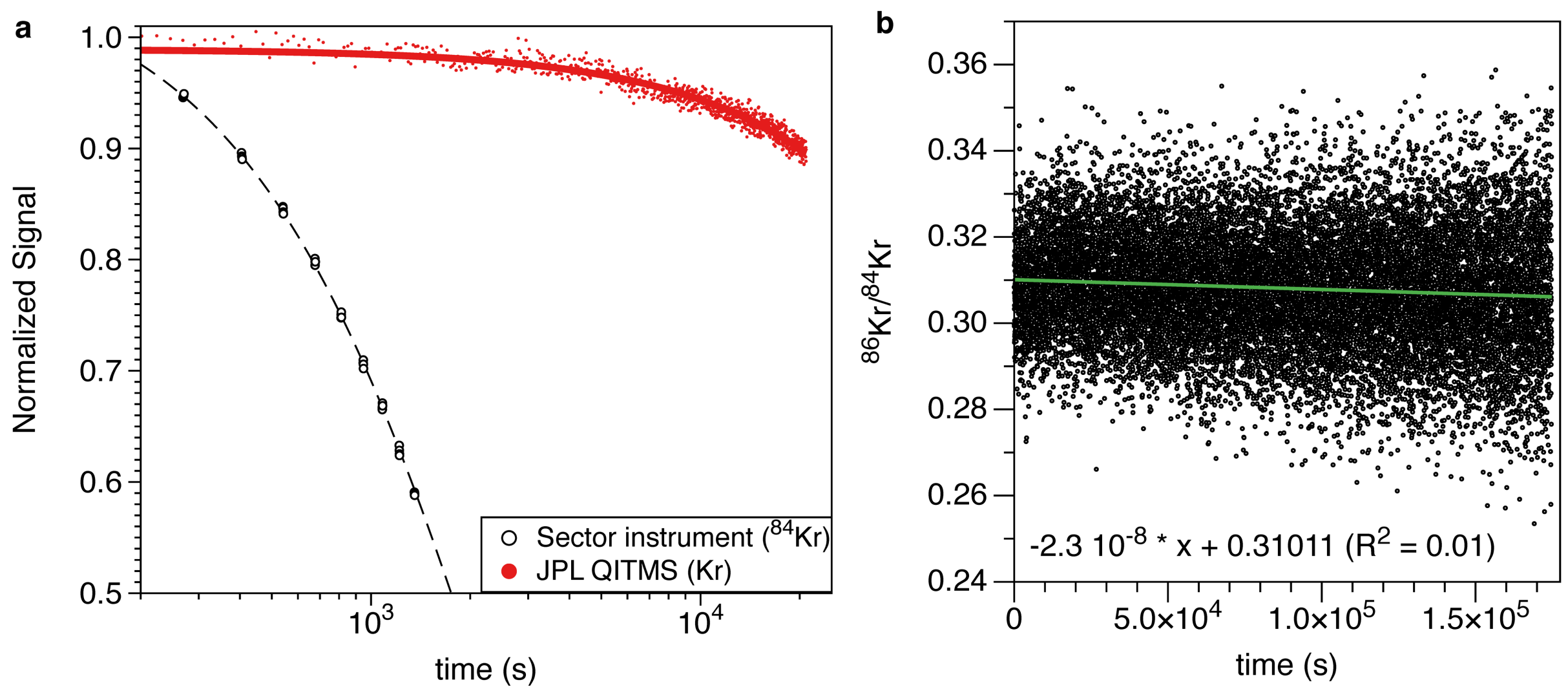

**Figure 2: Kr signal and $^{86}Kr/^{84}Kr$ ratio with time.** (a) The fast decrease of the Kr signal for the sector instrument restricts measurement of the gas to only about 30 min. The QITMS is less sensitive but the slow decrease in signal allows measurement for extended periods of time (1-24 h), improving counting statistics since numerous isotope ratios are collected. (b) Collected $^{86}Kr/^{84}Kr$ ratios with time. A linear regression through the dataset (green line) shows a slight decrease of the $^{86}Kr/^{84}Kr$ ratio with time (0.31011 (±0.0002) - 2.3*$10^{-8}$ (±2*$10^{-9}$) * time (s) , $R^2$ = 0.01).

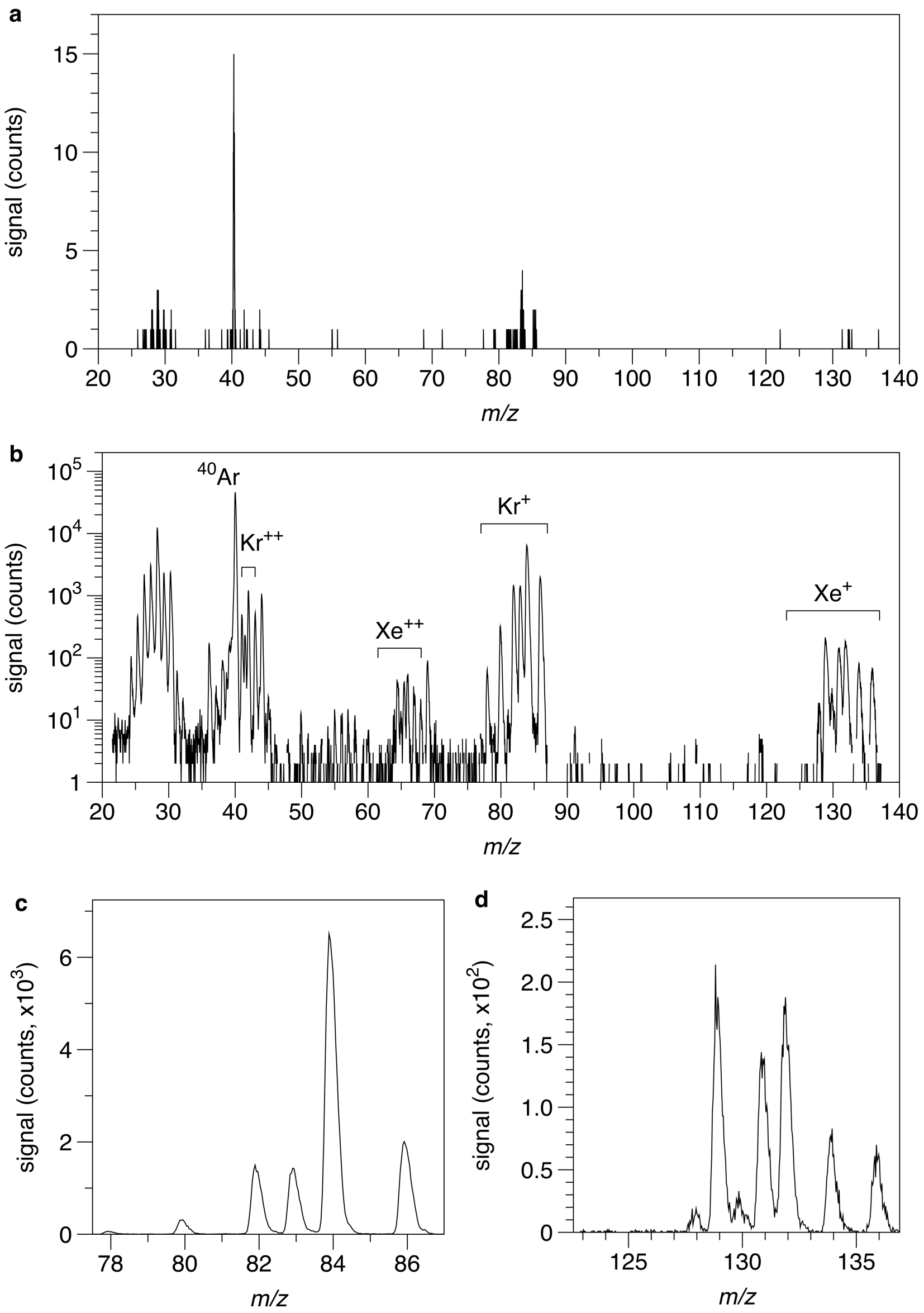

**Figure 3: Isotope spectra of an aliquot of calibrating gas obtained with the QITMS.** (a) Single frame obtained by the QITMS over 50 ms. Signals from $^{40}$Ar, Kr and Xe are visible. (b) Isotope spectrum from m/z = 20 to 137 obtained after 1 min of data acquisition. Contribution from singly and doubly charged Kr and Xe ions are visible as well as minor background contributions and Ar contamination. Note the logarithmic scale. (c) Isotope spectrum of krypton. (d) Isotope spectrum of xenon. $^{124}$Xe and $^{126}$Xe peaks are not visible on this scale.

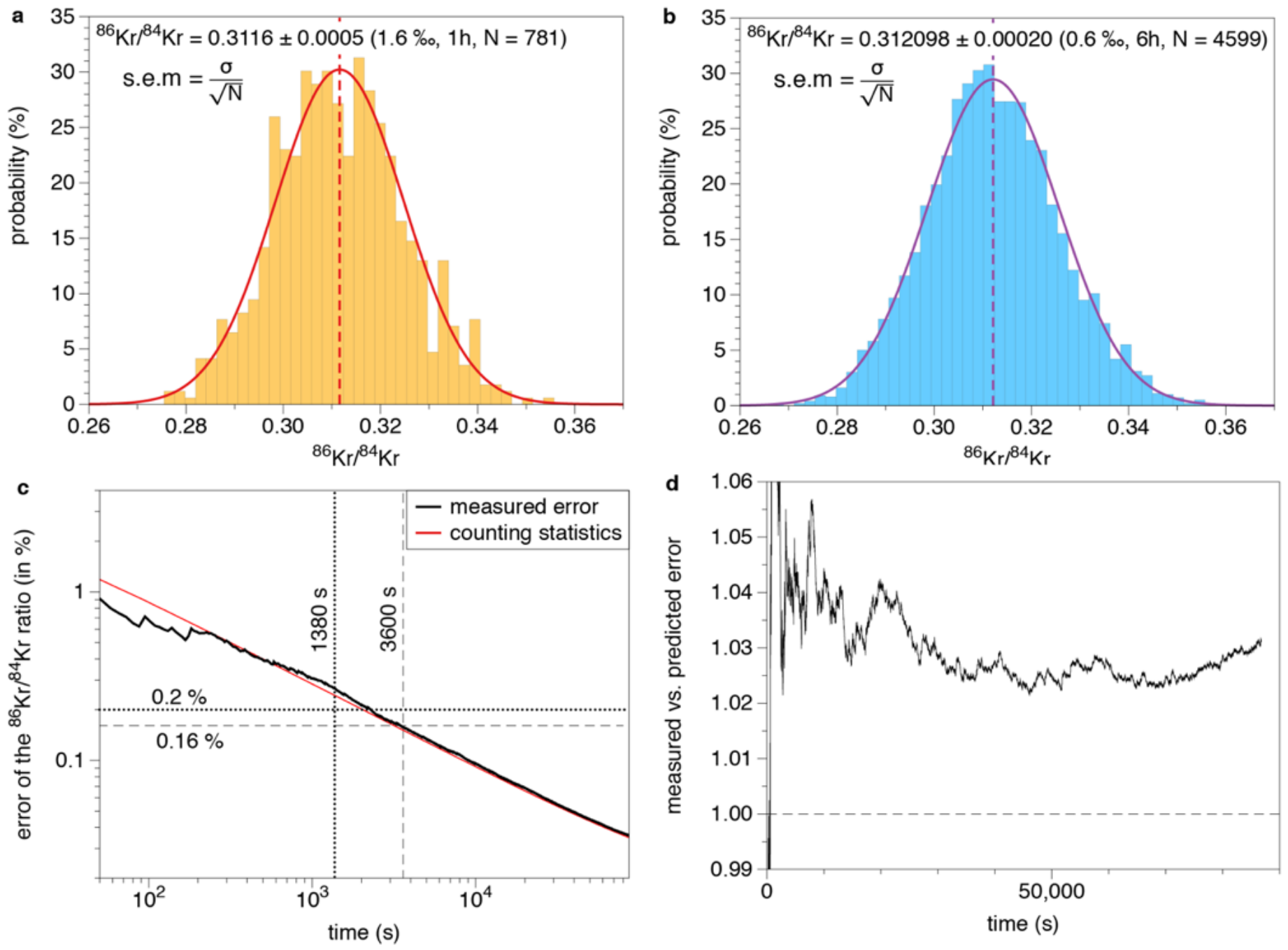

**Figure 4: Internal reproducibility of the instrument.** (a) Distribution of the $^{86}Kr/^{84}Kr$ ratios (N=781) for 1h of static measurement of an aliquot of the standard gas. Ratios are distributed following a normal distribution (red curve) centered at 0.3116 (dashed line). The standard deviation of the dataset ($\sigma$) is 0.015. The standard error of the mean is $\sigma/N^{1/2}$ = 0.0005 (1.6 ‰). Data have been bined with a bin size of 0.002. (b) Distribution of the $^{86}Kr/^{84}Kr$ ratios (N=4599) for 6h of static measurement on another aliquot of the standard gas. The precision is higher and reached 0.6 ‰. Data have been bined with a bin size of 0.002. (c) Evolution of the error on the $^{86}Kr/^{84}Kr$ ratio with the accumulation of isotope ratios for a 24h measurement. The measured error closely follows the error predicted when only counting statistics (Poisson's law) is considered (red curve). The black dashed lines correspond to the case presented in (a). The dotted line offers a comparison with the performance of a magnetic sector instrument. (d) Measured error divided by the error predicted by counting statistics with time. The measured error is only few percent higher than the predicted error. Errors at 1$\sigma$.

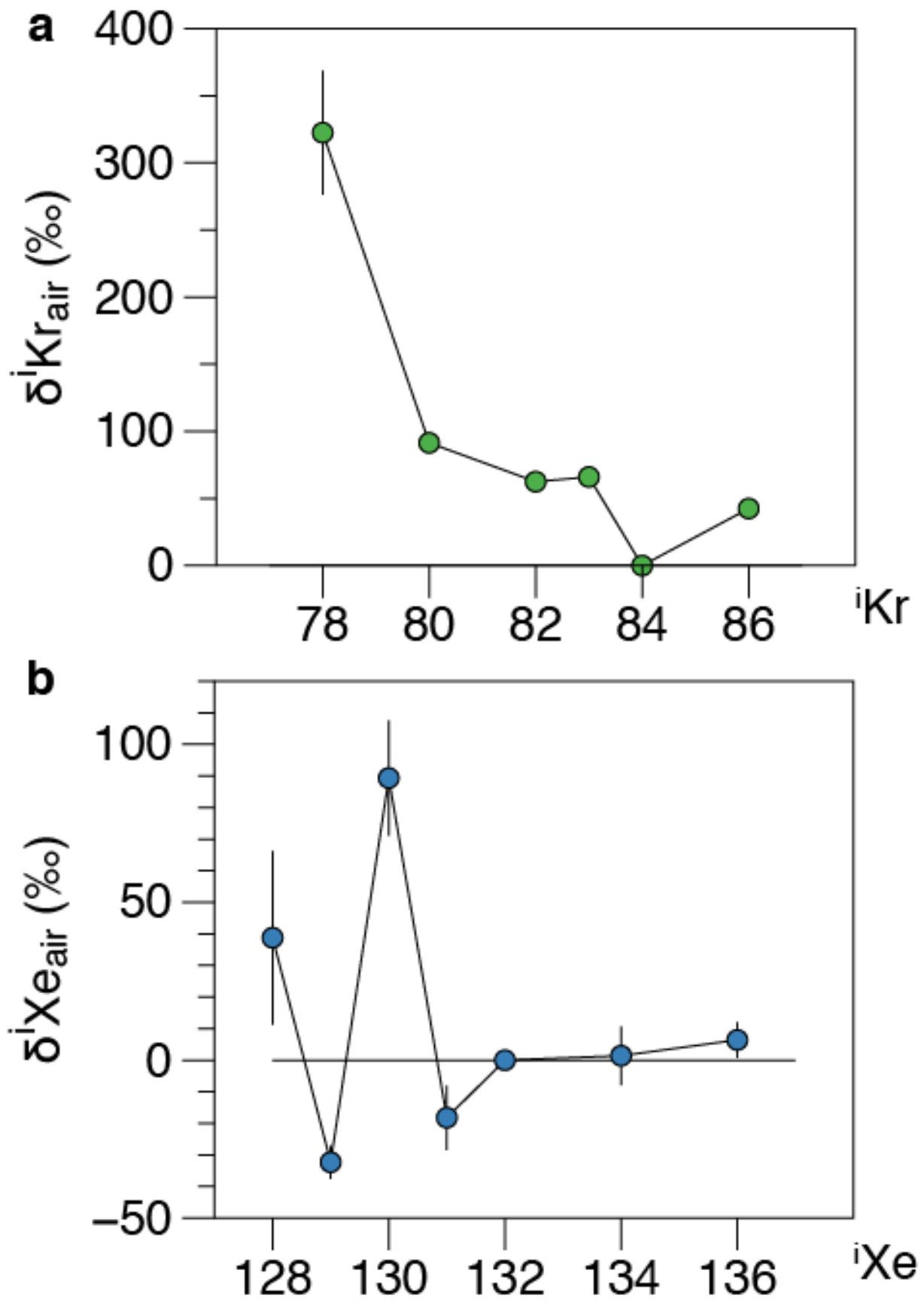

**Figure 5: Isotope ratios of Kr (a) and Xe (b) measured with the QITMS during the 1h experiments.** Isotope ratios are expressed relative to the isotopic composition of the atmosphere[24] and using the delta notation ($\delta^iXe_{air}$=(($^iXe/^{132}Xe)_{QITMS}$/($^iXe/^{132}Xe)_{air}$-1)×1,000). Mass discrimination effects are significant and justify the use of a standard gas in order to compute accurate isotope ratios after re-normalization. Errors at 1σ.

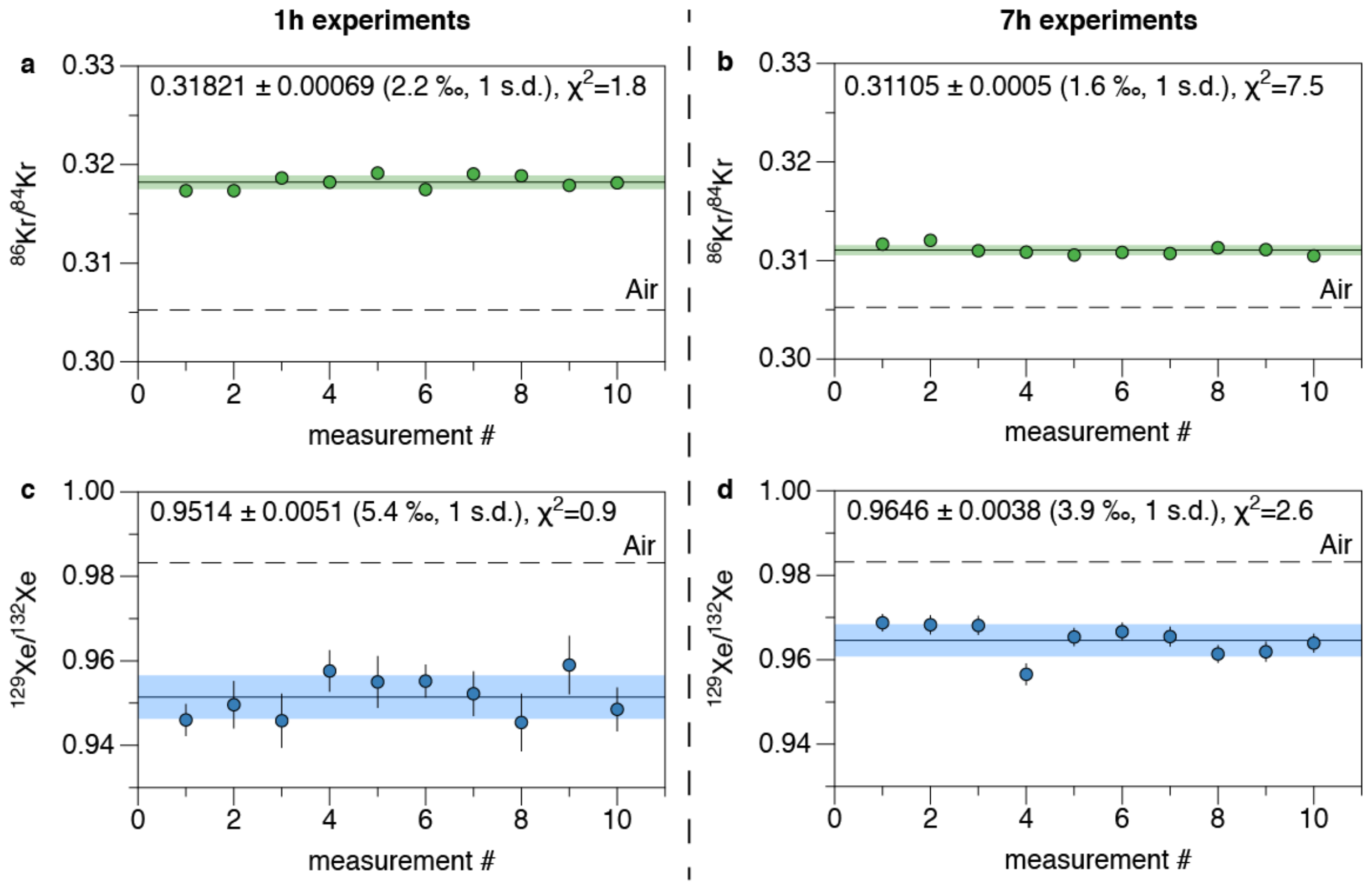

**Fig. 6: External reproducibility of the instrument for $^{86}Kr/^{84}Kr$ and $^{129}Xe/^{132}Xe$ ratios.** The plain lines depict the average value and the green range (for Kr) and the blue range (for Xe) correspond to the 1σ error range. (a) $^{86}Kr/^{84}Kr$ ratios for 10 measurements of about 1h. The dashed line corresponds to the isotopic ratio of atmospheric Kr (0.30524, ref. 26). (b) $^{86}Kr/^{84}Kr$ ratios for 10 measurements of about 7h. (c) $^{129}Xe/^{132}Xe$ ratios for 10 measurements of about 1h. The dashed line corresponds to the isotopic ratio of atmospheric Xe (0.9832, ref. 26). (d) $^{129}Xe/^{132}Xe$ ratios for 10 measurements of about 7h. Errors for individual measurements are given by $\sigma/N^{1/2}$ (see text) and for the dataset by the standard deviation of the different measurements. Errors at 1σ.

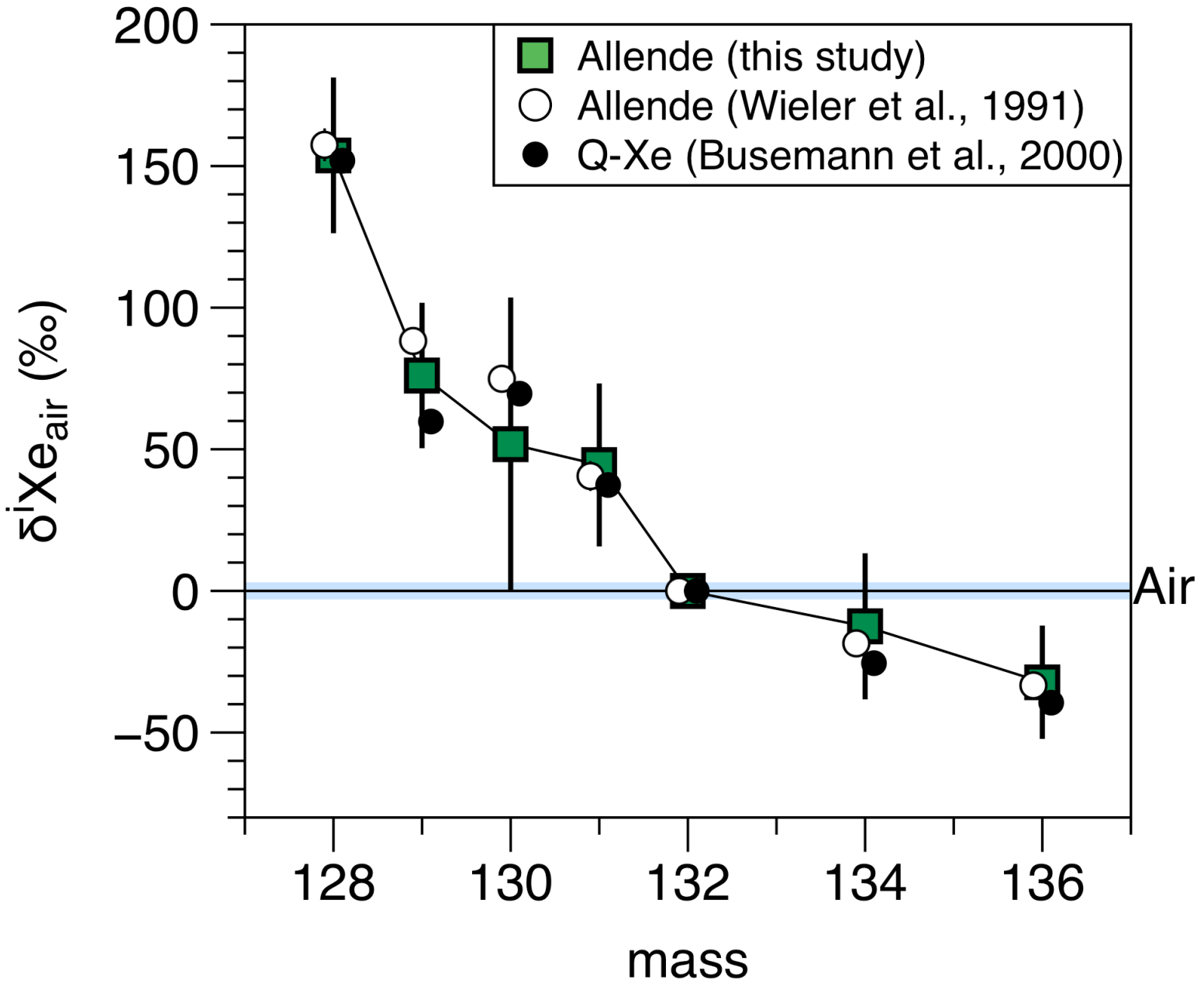

**Fig. 7: Isotopic composition of Xe extracted from Allende carbonaceous residue.** Isotope ratios are expressed relative to the isotopic composition of the atmosphere[24] and using the delta notation ($\delta^{i}Xe_{air}=((^{i}Xe/^{132}Xe)_{Allende}/(^{i}Xe/^{132}Xe)_{air}-1)\times1,000$). The isotope compositions of Xe of phase Q (ref. 21) and of a previous study on Allende (ref. 20) are shown for comparison. Error bars for literature data are not visible at this scale. Errors at 1σ.

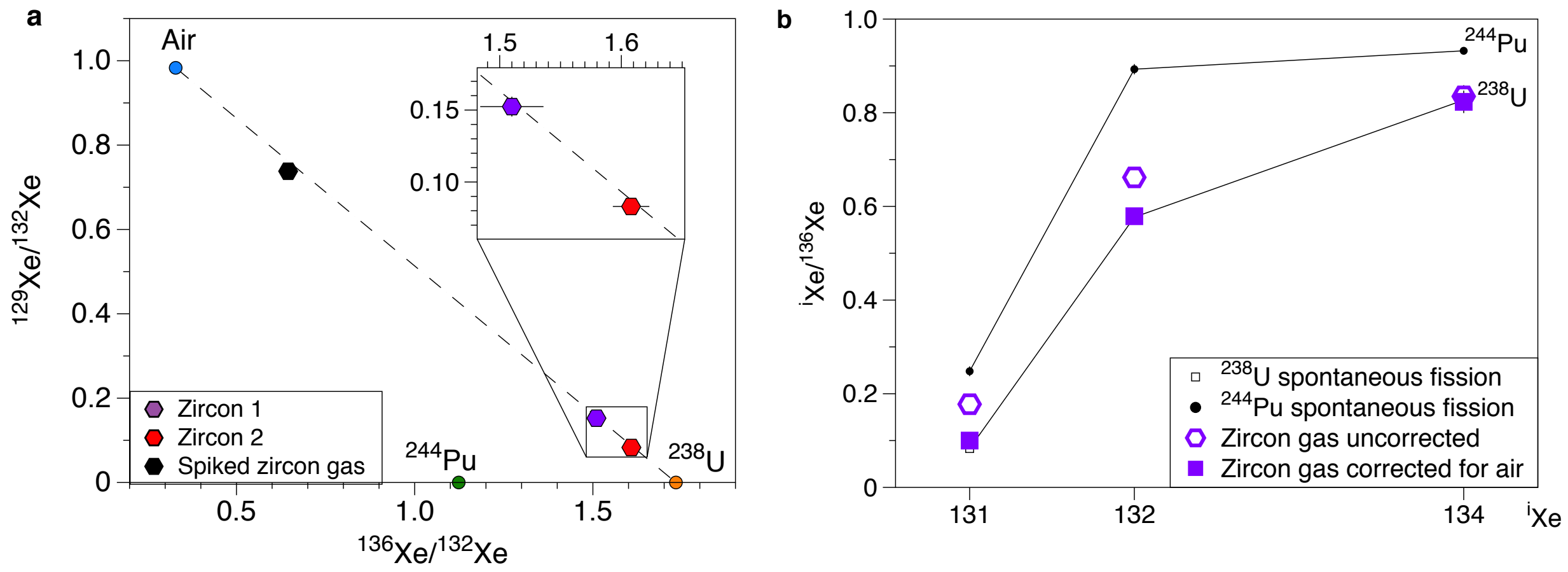

**Fig. 8: Isotopic composition of Xe measured in Sri Lanka zircon.** (a) Three isotope diagram. In this space, mixing between two component plots along lines. Pure $^{244}$Pu and $^{238}$U fission end-members are taken from ref. 23. The isotopic composition of Xe in air is from ref. 24. Data for zircons plot along the mixing line (dashed line) between air and the pure $^{238}$U fission end-members. "Spiked zircon gas" corresponds to the Zircon 2 gas mixed with an aliquot of the Kr/Xe standard gas of atmospheric composition. Data have been corrected for instrument discrimination. (b) Fission spectrum measured for Zircon 1 gas before (open symbols) and after (filled symbols) correction for air contamination. Isotope ratios are normalized to $^{136}$Xe, the major isotope produced by fission. Errors at 1σ.

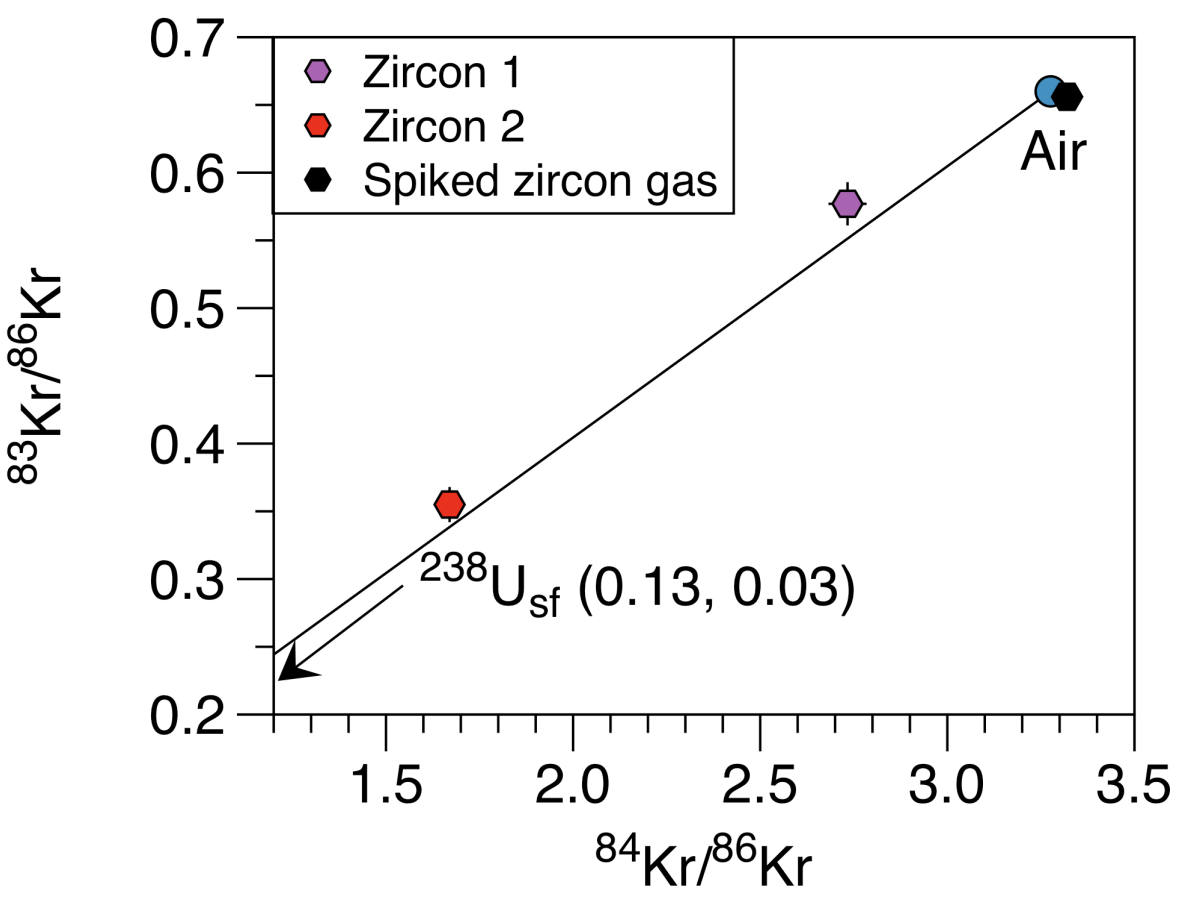

**Fig. 9: Three-isotope diagram showing the isotopic composition of Kr measured in zircons.**

There is a clear contribution either from the spontaneous fission of $^{238}$U ($^{238}U_{sf}$) for the zircon 1 and zircon 2 measurements. Errors at 1σ.

## Tables

**Table 1: Relevant experimental conditions.**

| Experiment duration | 1h | 7h-24h | 48h |
|---|---|---|---|
| Filament Current (A) | 1.65 | 1.65 | 1.65 |
| Energy (V) | -55 | -71.98 | -71.9 |
| Focus (V) | -46.15 | -62.53 | -62.53 |
| Frequency (Hz) | 1.06 | 1.06 | 1.06 |
| Multiplier Voltage (kV) | 2.72 | 2.72 | 2.72 |

**Table 2: Isotopic composition of Kr and Xe measured for aliquots of the calibrating gas measured during 1h. The isotopic compositions of atmospheric Kr and Xe are also listed as reference values.** Errors are at 1σ.

| Experiment ID | $^{84}$Kr (cps) | ± | Sensitivity (cps torr$^{-1}$) | | $^{78}$Kr/$^{84}$Kr | +/- | $^{80}$Kr/$^{84}$Kr | +/- | $^{82}$Kr/$^{84}$Kr | +/- | $^{83}$Kr/$^{84}$Kr | +/- | $^{86}$Kr/$^{84}$Kr | +/- |
|---|---|---|---|---|---|---|---|---|---|---|---|---|---|---|
| 2018-02-14-205g | 1101.3 | 0.5 | 1.126E+13 | 5E+09 | 0.00869 | 0.00007 | 0.04329 | 0.00014 | 0.21548 | 0.00034 | 0.21461 | 0.00035 | 0.31735 | 0.00042 |
| 2018-02-14-206g | 1087.7 | 2.1 | 1.113E+13 | 2E+10 | 0.00810 | 0.00007 | 0.04288 | 0.00018 | 0.21385 | 0.00043 | 0.21365 | 0.00042 | 0.31735 | 0.00054 |
| 2018-02-14-207g | 1085.3 | 2.7 | 1.111E+13 | 3E+10 | 0.00826 | 0.00009 | 0.04312 | 0.00021 | 0.21423 | 0.00050 | 0.21398 | 0.00052 | 0.31864 | 0.00061 |
| 2018-02-15-201g | 1135.4 | 2.2 | 1.162E+13 | 2E+10 | 0.00788 | 0.00007 | 0.04341 | 0.00017 | 0.21395 | 0.00041 | 0.21478 | 0.00043 | 0.31822 | 0.00050 |
| 2018-02-15-202g | 1116.9 | 2.2 | 1.143E+13 | 2E+10 | 0.00781 | 0.00007 | 0.04336 | 0.00016 | 0.21512 | 0.00038 | 0.21555 | 0.00039 | 0.31914 | 0.00048 |
| 2018-02-15-203g | 1121.3 | 3.1 | 1.148E+13 | 3E+10 | 0.00800 | 0.00008 | 0.04338 | 0.00018 | 0.21567 | 0.00045 | 0.21480 | 0.00043 | 0.31747 | 0.00054 |
| 2018-02-15-204g | 1118.3 | 2.6 | 1.145E+13 | 3E+10 | 0.00787 | 0.00008 | 0.04350 | 0.00019 | 0.21516 | 0.00045 | 0.21443 | 0.00046 | 0.31905 | 0.00057 |
| 2018-02-15-205g | 1110 | 3.8 | 1.137E+13 | 4E+10 | 0.00819 | 0.00009 | 0.04343 | 0.00020 | 0.21538 | 0.00046 | 0.21518 | 0.00048 | 0.31886 | 0.00059 |
| 2018-02-15-206g | 1109.9 | 3.0 | 1.137E+13 | 3E+10 | 0.00797 | 0.00008 | 0.04289 | 0.00020 | 0.21436 | 0.00048 | 0.21458 | 0.00046 | 0.31789 | 0.00060 |
| 2018-02-16-201g | 1161.1 | 2.5 | 1.190E+13 | 3E+10 | 0.00773 | 0.00007 | 0.04294 | 0.00016 | 0.21495 | 0.00040 | 0.21498 | 0.00041 | 0.31815 | 0.00049 |
| **Average** | | | **1.141E+13** | **2.3E+11** | **0.00805** | **0.00028** | **0.04322** | **0.00024** | **0.21481** | **0.00066** | **0.21465** | **0.00055** | **0.31821** | **0.00069** |
| **MSWD - reduced $\chi^2$** | | | **177** | | **17.0** | | **1.8** | | **2.4** | | **1.7** | | **1.8** | |
| **Air** | | | | | **0.00609** | **0.00002** | **0.03960** | **0.00020** | **0.20217** | **0.00004** | **0.20136** | **0.00021** | **0.30524** | **0.00025** |

| Experiment ID | $^{132}$Xe (cps) | ± | Sensitivity (cps torr$^{-1}$) | | $^{128}$Xe/$^{132}$Xe | +/- | $^{129}$Xe/$^{132}$Xe | +/- | $^{130}$Xe/$^{132}$Xe | +/- | $^{131}$Xe/$^{132}$Xe | +/- | $^{134}$Xe/$^{132}$Xe | +/- | $^{136}$Xe/$^{132}$Xe | +/- |
|---|---|---|---|---|---|---|---|---|---|---|---|---|---|---|---|---|
| 2018-02-14-205g | 46.36 | 0.34 | 1.39E+13 | 1E+11 | 0.0735 | 0.0010 | 0.9460 | 0.0038 | 0.1632 | 0.0013 | 0.7709 | 0.0031 | 0.3872 | 0.0025 | 0.3292 | 0.0022 |
| 2018-02-14-206g | 45.04 | 0.49 | 1.35E+13 | 1E+11 | 0.0735 | 0.0012 | 0.9496 | 0.0056 | 0.1656 | 0.0018 | 0.7810 | 0.0034 | 0.3919 | 0.0029 | 0.3302 | 0.0023 |
| 2018-02-14-207g | 46.7 | 0.7 | 1.40E+13 | 2E+11 | 0.0707 | 0.0012 | 0.9458 | 0.0064 | 0.1615 | 0.0021 | 0.7773 | 0.0041 | 0.3888 | 0.0030 | 0.3313 | 0.0035 |
| 2018-02-15-201g | 47.26 | 0.46 | 1.41E+13 | 1E+11 | 0.0749 | 0.0011 | 0.9576 | 0.0049 | 0.1679 | 0.0019 | 0.7766 | 0.0048 | 0.3923 | 0.0026 | 0.3334 | 0.0024 |
| 2018-02-15-202g | 46.45 | 0.45 | 1.39E+13 | 1E+11 | 0.0726 | 0.0013 | 0.9550 | 0.0061 | 0.1637 | 0.0016 | 0.7747 | 0.0047 | 0.3855 | 0.0025 | 0.3322 | 0.0027 |
| 2018-02-15-203g | 50.87 | 0.69 | 1.52E+13 | 2E+11 | 0.0748 | 0.0018 | 0.9552 | 0.0039 | 0.1659 | 0.0019 | 0.7779 | 0.0051 | 0.3842 | 0.0030 | 0.3306 | 0.0030 |
| 2018-02-15-204g | 47.47 | 0.58 | 1.42E+13 | 2E+11 | 0.0781 | 0.0014 | 0.9522 | 0.0053 | 0.1681 | 0.0019 | 0.7826 | 0.0063 | 0.3933 | 0.0030 | 0.3345 | 0.0027 |
| 2018-02-15-205g | 47.24 | 0.63 | 1.41E+13 | 2E+11 | 0.0744 | 0.0021 | 0.9454 | 0.0068 | 0.1630 | 0.0025 | 0.7570 | 0.0045 | 0.3906 | 0.0034 | 0.3298 | 0.0032 |
| 2018-02-15-206g | 47.2 | 0.66 | 1.41E+13 | 2E+11 | 0.0755 | 0.0014 | 0.9590 | 0.0069 | 0.1688 | 0.0025 | 0.7820 | 0.0052 | 0.3876 | 0.0036 | 0.3335 | 0.0030 |
| 2018-02-16-201g | 48.85 | 0.64 | 1.46E+13 | 2E+11 | 0.0732 | 0.0009 | 0.9485 | 0.0052 | 0.1612 | 0.0015 | 0.7664 | 0.0044 | 0.3831 | 0.0030 | 0.3304 | 0.0023 |
| **Average** | | | **1.42E+13** | **5E+11** | **0.0741** | **0.0020** | **0.9514** | **0.0051** | **0.1649** | **0.0028** | **0.7746** | **0.0080** | **0.3884** | **0.0035** | **0.3315** | **0.0018** |
| **MSWD - reduced $\chi^2$** | | | **7.4** | | **2.2** | | **0.9** | | **2.1** | | **3.2** | | **1.5** | | **0.5** | |
| **Air** | | | | | **0.07136** | **0.00009** | **0.9832** | **0.0012** | **0.15136** | **0.00012** | **0.789** | **0.0011** | **0.3879** | **0.0006** | **0.3294** | **0.0004** |

**Table 3: Isotopic composition of Kr and Xe measured for aliquots of the calibrating gas measured during 6-8h.** Errors are at 1σ.

| periment ID | $^{84}$Kr (cps) | ± | Sensitivity (cps torr$^{-1}$) | | $^{78}$Kr/$^{84}$Kr | +/- | $^{80}$Kr/$^{84}$Kr | +/- | $^{82}$Kr/$^{84}$Kr | +/- | $^{83}$Kr/$^{84}$Kr | +/- | $^{86}$Kr/$^{84}$Kr | +/- |
|---|---|---|---|---|---|---|---|---|---|---|---|---|---|---|
| 4g | 1272 | 0.7 | 1.3110E+13 | 7E+09 | 0.007102 | 0.000025 | 0.04264 | 0.00006 | 0.21305 | 0.00015 | 0.21191 | 0.00015 | 0.31166 | 0.00019 |
| 2a | 1270.4 | 0.7 | 1.3099E+13 | 7E+09 | 0.007061 | 0.000026 | 0.04261 | 0.00006 | 0.21273 | 0.00015 | 0.21198 | 0.00015 | 0.31205 | 0.00019 |
| 2g | 1279.3 | 0.7 | 1.3194E+13 | 7E+09 | 0.007061 | 0.000025 | 0.04259 | 0.00006 | 0.21244 | 0.00015 | 0.21169 | 0.00015 | 0.31099 | 0.00019 |
| 3a | 1274.4 | 0.8 | 1.3146E+13 | 8E+09 | 0.007012 | 0.000027 | 0.04282 | 0.00007 | 0.21343 | 0.00016 | 0.21167 | 0.00016 | 0.31085 | 0.00020 |
| 2a | 1291 | 0.6 | 1.3320E+13 | 6E+09 | 0.006983 | 0.000024 | 0.04286 | 0.00006 | 0.21354 | 0.00014 | 0.21231 | 0.00014 | 0.31055 | 0.00018 |
| 2a | 1267.3 | 0.7 | 1.3078E+13 | 7E+09 | 0.006974 | 0.000025 | 0.04291 | 0.00006 | 0.21340 | 0.00015 | 0.21248 | 0.00015 | 0.31082 | 0.00019 |
| 2a | 1267.3 | 0.7 | 1.3081E+13 | 7E+09 | 0.006980 | 0.000026 | 0.04282 | 0.00007 | 0.21383 | 0.00016 | 0.21269 | 0.00015 | 0.31071 | 0.00020 |
| 2a | 1269.4 | 0.7 | 1.3106E+13 | 7E+09 | 0.006985 | 0.000027 | 0.04296 | 0.00007 | 0.21368 | 0.00016 | 0.21259 | 0.00016 | 0.31130 | 0.00020 |
| 2a | 1270.6 | 0.8 | 1.3121E+13 | 8E+09 | 0.007012 | 0.000028 | 0.04290 | 0.00007 | 0.21341 | 0.00017 | 0.21186 | 0.00017 | 0.31110 | 0.00021 |
| 1g | 1289 | 0.7 | 1.3314E+13 | 8E+09 | 0.006904 | 0.000028 | 0.04266 | 0.00007 | 0.21309 | 0.00017 | 0.21204 | 0.00017 | 0.31047 | 0.00021 |
| | | | **1.316E+13** | **9.08E+10** | **0.007007** | **0.000056** | **0.04278** | **0.00014** | **0.21326** | **0.00043** | **0.21212** | **0.00037** | **0.31105** | **0.00050** |
| **ced χ$^2$** | | | **173** | | | **4.6** | | **4.5** | | **7.9** | | **5.8** | | **7.5** |

| periment ID | $^{132}$Xe (cps) | ± | Sensitivity (cps torr$^{-1}$) | | $^{128}$Xe/$^{132}$Xe | +/- | $^{129}$Xe/$^{132}$Xe | +/- | $^{130}$Xe/$^{132}$Xe | +/- | $^{131}$Xe/$^{132}$Xe | +/- | $^{134}$Xe/$^{132}$Xe | +/- | $^{136}$Xe/$^{132}$Xe | +/- | | | | |
|---|---|---|---|---|---|---|---|---|---|---|---|---|---|---|---|---|---|---|---|---|
| 4g | 53.28 | 0.14 | 1.605E+13 | 4E+10 | 0.00396 | 0.00010 | 0.00410 | 0.00011 | 0.07325 | 0.00041 | 0.9688 | 0.0021 | 0.1772 | 0.0007 | 0.7869 | 0.0018 | 0.3823 | 0.0012 | 0.3247 | 0.0010 |
| 2a | 51.44 | 0.16 | 1.551E+13 | 5E+10 | 0.00508 | 0.00011 | 0.00520 | 0.00011 | 0.07430 | 0.00047 | 0.9683 | 0.0023 | 0.1780 | 0.0007 | 0.7844 | 0.0020 | 0.3864 | 0.0012 | 0.3247 | 0.0010 |
| 2g | 50.27 | 0.18 | 1.516E+13 | 5E+10 | 0.00383 | 0.00010 | 0.00411 | 0.00010 | 0.07358 | 0.00044 | 0.9681 | 0.0023 | 0.1761 | 0.0007 | 0.7844 | 0.0020 | 0.3819 | 0.0012 | 0.3228 | 0.0011 |
| 3a | 56.45 | 0.21 | 1.702E+13 | 6E+10 | 0.00395 | 0.00010 | 0.00404 | 0.00011 | 0.07448 | 0.00066 | 0.9566 | 0.0026 | 0.1778 | 0.0013 | 0.7799 | 0.0020 | 0.3807 | 0.0013 | 0.3207 | 0.0011 |
| 2a | 52.48 | 0.15 | 1.583E+13 | 5E+10 | 0.00362 | 0.00008 | 0.00353 | 0.00009 | 0.07501 | 0.00042 | 0.9654 | 0.0021 | 0.1762 | 0.0007 | 0.7799 | 0.0017 | 0.3804 | 0.0010 | 0.3233 | 0.0009 |
| 2a | 51.20 | 0.17 | 1.545E+13 | 5E+10 | 0.00358 | 0.00010 | 0.00371 | 0.00009 | 0.07349 | 0.00045 | 0.9666 | 0.0022 | 0.1779 | 0.0007 | 0.7814 | 0.0018 | 0.3834 | 0.0011 | 0.3237 | 0.0010 |
| 2a | 47.48 | 0.19 | 1.433E+13 | 6E+10 | 0.00371 | 0.00009 | 0.00370 | 0.00009 | 0.07462 | 0.00047 | 0.9655 | 0.0023 | 0.1766 | 0.0007 | 0.7853 | 0.0019 | 0.3835 | 0.0012 | 0.3256 | 0.0010 |
| 2a | 49.75 | 0.18 | 1.502E+13 | 5E+10 | 0.00403 | 0.00010 | 0.00375 | 0.00009 | 0.07534 | 0.00046 | 0.9614 | 0.0021 | 0.1764 | 0.0008 | 0.7816 | 0.0018 | 0.3827 | 0.0011 | 0.3238 | 0.0010 |
| 2a | 50.21 | 0.22 | 1.516E+13 | 7E+10 | 0.00370 | 0.00010 | 0.00365 | 0.00009 | 0.07435 | 0.00046 | 0.9619 | 0.0024 | 0.1756 | 0.0008 | 0.7808 | 0.0019 | 0.3811 | 0.0012 | 0.3233 | 0.0011 |
| 1g | 51.38 | 0.22 | 1.551E+13 | 7E+10 | 0.00397 | 0.00011 | 0.00429 | 0.00012 | 0.07329 | 0.00043 | 0.9639 | 0.0022 | 0.1775 | 0.0008 | 0.7834 | 0.0020 | 0.3844 | 0.0013 | 0.3289 | 0.0011 |
| | | | **1.55E+13** | **7E+11** | **0.00394** | **0.00043** | **0.00401** | **0.00049** | **0.07417** | **0.00073** | **0.9646** | **0.0038** | **0.1769** | **0.0009** | **0.7828** | **0.0024** | **0.3827** | **0.0018** | **0.3242** | **0.0021** |
| **ced χ$^2$** | | | **152** | | | **15.3** | | **23.2** | | **2.8** | | **2.6** | | **1.2** | | **1.7** | | **2.4** | | **3.9** |

**Table 4: Isotopic composition of Kr and Xe measured for gas extracted from chemical residues of Allende and from a 600 Myr-old zircon megacryst.** Spiked zircon gas corresponds to a mixture of the ZIR2 gas and an aliquot of the Kr-Xe standard. All data have been corrected for instrument effects. Errors are at 1$\sigma$.

| **Experiment ID** | **$^{78}Kr/^{84}Kr$** | +/- | **$^{80}Kr/^{84}Kr$** | +/- | **$^{82}Kr/^{84}Kr$** | +/- | **$^{83}Kr/^{84}Kr$** | +/- | **$^{86}Kr/^{84}Kr$** | +/- |
|---|---|---|---|---|---|---|---|---|---|---|
| **Zircons** | | | | | | | | | | |
| ZIR1 | n.d. | | 0.044 | 0.002 | 0.208 | 0.002 | 0.211 | 0.004 | 0.366 | 0.007 |
| ZIR2 | n.d. | | 0.039 | 0.002 | 0.182 | 0.008 | 0.213 | 0.006 | 0.599 | 0.015 |
| **Allende** | | | | | | | | | | |
| Pt1 | n.d. | | n.d. | | 0.2046 | 0.0062 | 0.2214 | 0.0073 | 0.2865 | 0.0078 |
| Pt2 | n.d. | | n.d. | | 0.1925 | 0.0053 | 0.2011 | 0.0063 | 0.2909 | 0.0076 |
| Pt3 | n.d. | | n.d. | | 0.1942 | 0.0023 | 0.2030 | 0.0021 | 0.2930 | 0.0033 |
| Pt4 | n.d. | | n.d. | | 0.2006 | 0.0019 | 0.1991 | 0.0020 | 0.2990 | 0.0026 |
| **Average** | | | | | **0.1980** | **0.0056** | **0.2062** | **0.0103** | **0.2924** | **0.0052** |
| **reduced $\chi^2$ - MSWD** | | | | | | **2.2** | | **6.7** | | **2.4** |

| **Experiment ID** | **$^{128}Xe/^{132}Xe$** | +/- | **$^{129}Xe/^{132}Xe$** | +/- | **$^{130}Xe/^{132}Xe$** | +/- | **$^{131}Xe/^{132}Xe$** | +/- | **$^{134}Xe/^{132}Xe$** | +/- | **$^{136}Xe/^{132}Xe$** | +/- |
|---|---|---|---|---|---|---|---|---|---|---|---|---|
| **Zircons** | | | | | | | | | | | | |
| ZIR1 | n.d. | | 0.152 | 0.0053 | n.d. | | 0.2681 | 0.0068 | 1.264 | 0.018 | 1.518 | 0.025 |
| ZIR2 | n.d. | | 0.083 | 0.002 | n.d. | | 0.210 | 0.004 | 1.342 | 0.016 | 1.608 | 0.015 |
| Spiked zircon gas | 0.050800 | 0.00180 | 0.738 | 0.007 | 0.116 | 0.002 | 0.629 | 0.009 | 0.626 | 0.007 | 0.646 | 0.006 |
| **Allende** | | | | | | | | | | | | |
| Pt1 | 0.0831 | 0.0034 | 1.0815 | 0.0193 | 0.1548 | 0.0052 | 0.8493 | 0.0164 | 0.3931 | 0.0089 | 0.3267 | 0.0082 |
| Pt2 | 0.0847 | 0.0037 | 1.0779 | 0.0218 | 0.1524 | 0.0051 | 0.8369 | 0.0152 | 0.3892 | 0.0076 | 0.3133 | 0.0068 |
| Pt3 | 0.0802 | 0.0026 | 1.0389 | 0.0084 | 0.1596 | 0.0033 | 0.8084 | 0.0100 | 0.3711 | 0.0048 | 0.3135 | 0.0033 |
| Pt4 | 0.0814 | 0.0023 | 1.0336 | 0.0067 | 0.1701 | 0.0031 | 0.8018 | 0.0089 | 0.3788 | 0.0039 | 0.3217 | 0.0026 |
| **Average** | **0.0823** | **0.0020** | **1.0580** | **0.0252** | **0.1592** | **0.0078** | **0.8241** | **0.0227** | **0.3830** | **0.0100** | **0.3188** | **0.0066** |
| **reduced $\chi^2$ - MSWD** | | **0.4** | | **6.9** | | **5.0** | | **3.9** | | **3.1** | | **1.8** |